\documentclass{aa}
\usepackage{txfonts}
\usepackage{epsfig}
\usepackage{graphicx}
\usepackage{amssymb}	
\usepackage{natbib}
\usepackage{latexsym}
\usepackage{hyperref}
\usepackage[small,hang,bf]{caption}
\bibpunct{(}{)}{;}{}{}{,}

\usepackage{graphicx}
\usepackage{xspace}
\newcommand{\ang}{ \AA\xspace}

\newcommand{\Nh}{\rm $N_{\rm\scriptsize{H}}$\xspace}
\newcommand{\nh}{\rm $n_{\rm\scriptsize{H}}$\xspace}
\newcommand{\nel}{\rm $n_{\rm\scriptsize{e}}$\xspace}

\def\Msun{\ensuremath{\mathrm{M_\odot}}\xspace}
\newcommand{\degree}{\ensuremath{^\circ}\xspace}

\def\apjl{ApJ}%
\def\apjs{ApJS}%
\def\ao{Appl.~Opt.}%
\title{The high resolution X-ray spectrum of SNR 0506-68 using XMM-Newton }
\author{S. Broersen
\and J. Vink
\and J. Kaastra
\and J. Raymond}

\institute{Astronomical Institute, University of Utrecht, Postbus 80000, 3508 TA Utrecht, The Netherlands \email{s.broersen@astro-uu.nl}\\
SRON, Netherlands Institute for Space Research, Sorbonnelaan 2, 3584 CA Utrecht, The Netherlands\\ 
Harvard-Smithsonian Center for Astrophysics, 60 Garden Street, Cambridge, MA 02138, USA \\
}
\date{Received 1 June 2011 / Accepted 30 August  2011}
\abstract{}{We study the supernova remnant 0506-68 in order to obtain detailed information about, among other things, the ionisation state and age of the ionised plasma.}{Using the Reflection Grating Spectrometer (RGS) onboard the XMM-Newton satellite we are able to take detailed spectra of the remnant. In addition, we use the MOS data to obtain spectral information at higher energies. }{The spectrum shows signs of recombination and we derive the conditions for which the remnant and SNR in general are able to cool rapidly enough to become over-ionised. The elemental abundances found are mostly in agreement with the mean LMC abundances. Our models and calculations favour the lower age estimate mentioned in the literature of $\sim4000$ year.  }{}
\titlerunning{The high-resolution X-ray spectrum of SNR 0506-68 using XMM-Newton.}
\authorrunning{S. Broersen et al.}
\begin{document}

\maketitle
\section{Introduction}

Supernova remnants (SNRs) hold important information about the nucleosynthesis and energy of the supernova explosion, its circumstellar matter (CSM) evolution and the physics of the ionised plasmas. The Large Magellanic Cloud (LMC) is particularly well suited for the study of SNRs, since the absorption column to the LMC is low and it is relatively close  (50\,kpc). Moreover, studying SNRs in the LMC has the advantage that the distance is known quite precisely, which is convenient for luminosity and length scale calculations.

\begin{figure}[h]
\centering
\resizebox{\hsize}{!}{\includegraphics{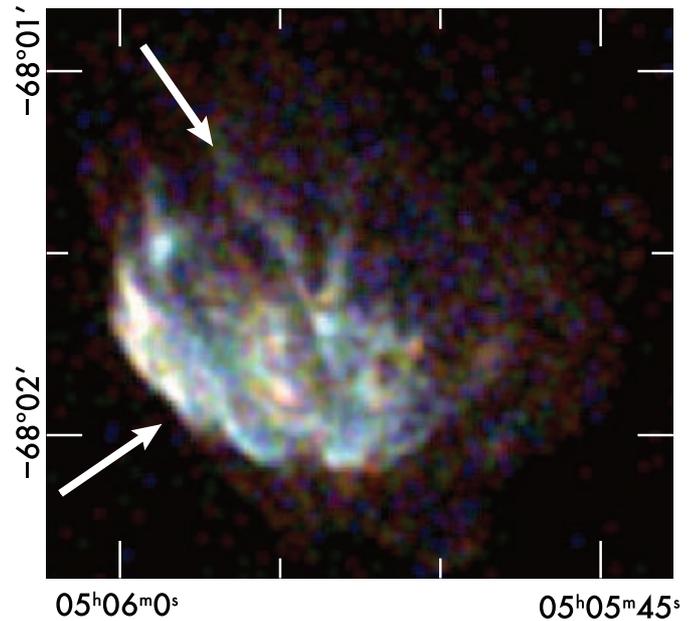}}
\caption{A smoothed Chandra RGB image of the SNR 0506-68. The arrows denote the direction of the RGS dispersion directions. The RGS field of view covers the remnant completely. Red corresponds to \ion{O}{vii} (0.53-0.61 keV), green to \ion{Fe}{L} (0.79-0.89 keV) and blue to \ion{Mg}{xi} (1.25-1.41 keV). The arrows denote the dispersion axis orientation of the 2000 (upper arrow) and the 2002 (lower arrow) observations.}  
\label{Fig.:rgb}
\end{figure}

SNR 0506-68 (also known as N23) is a small ($R \sim 10\,$pc, see Fig. \ref{Fig.:rgb}) remnant located fairly centrally in the LMC. The X-ray emission of the remnant is characterised by a filamentary structure with some bright spots. There is a gradient in brightness running from southeast to northwest. The remnant has been studied by \cite{hughesetal2006} with the Chandra telescope. Both \cite{hughesetal2006} and \cite{hayatoetal2006} report the presence of a compact object and conclude that the SNR is a result of a core-collapse supernova explosion (SNe). They find that the X-ray emission comes largely from the swept-up interstellar medium and estimate the age of the remnant to be $\sim$4600 yr. Recently, \cite{someyaetal2010} used the XIS instrument onboard the Suzaku telescope to obtain a more detailed spectrum of SNR 0506-68. They note the presence of a cool ($\sim$0.2 keV) temperature component in addition to a hot ($\sim$0.6 keV) component in the plasma, and a high ionisation parameter $n_{\rm e} t$ ($\sim$10$^{13}$ cm$^{-3}$ s). From a Sedov analysis they conclude that the age of the remnant, based on the cool component, may be as high as $\sim$8000 yr, but that it may have entered the radiative phase of its evolution. If the age is indeed as high as 8000 yr that would mean that the size of the remnant is quite small for its age, suggesting that the explosion took place in a high density region of the LMC. 

It has long been recognized that when a gas is suddenly heated in a shock it is \emph{under}-ionised, and requires a density-weighted time scale $n_{\rm e}t\sim10^{12}$ cm$^{-3}$ s to approach equilibrium \citep{smithhughes2010}. Recently, however, two groups have reported evidence for \emph{over}-ionised plasma in the SNRs W49B and IC 443 \citep{yamaguchietal2009,ozawaetal2009,miceli2010}. In this paper we aim to obtain detailed understanding about the ionised plasma of SNR 0506-68 using the spectral diagnostic capacity of the RGS (Reflection Grating Spectrometer) instrument \citep{denherderetal2001}. High resolution spectra of SNRs, obtained with grating spectrometers such as the RGS, hold detailed information about the ionised plasma. The spectral resolution of the RGS is large enough to resolve, among others, the OVII He-$\alpha$ triplet at $\sim22$\ang and the $\ion{Fe}{xvii}$ line complex around 15-17\ang, provided the source has a small angular extent. The ratios of different emission lines in these elements provide interesting plasma diagnostics \citep[e.g.:][]{porquetetal2011}.

\section{Data}
We used the data obtained by the XXM-Newton satellite on July 6, 2000 (obs ID 0111130101) and July 10, 2002 (obs ID 011130701).  The exposure times of the observations are 18.1 and 19.7 ks. For our spectral study, we made use of the RGS and EPIC MOS \citep{turneretal2001} data. Although the EPIC pn camera has a higher sensitivity, the MOS cameras have a higher spectral resolution, which is important for comparison with the RGS data. All MOS observations were performed in full frame mode with the medium filter in place. 

The 2000 RGS observation was taken with the dispersion direction oriented 27\degree counterclockwise from the celestial north, while the 2002 observation was rotated 90\degree with respect to this observation, at a dispersion direction orientation of 117\degree (see Fig. \ref{Fig.:rgb}). This resulted in different line profiles which need separate responses, as the 2002 observation has a higher effective resolution than the 2000 one. The RGS data were corrected for periods of high background flaring by creating good time intervals based on the count rate in CCD number 9 of the instrument. This CCD is closest to the optical axis of the telescope and therefore affected most by background flaring. The second order spectra have a higher resolution, but are of lower statistical quality. Since they provide no additional information they are not presented here. 

The RGS is a slitless spectrometer. When using this kind of instruments, the  lines in the spectra get smeared out as a result of the extent of the source. Although the angular size of SNRs observed in the LMC is modest, this smearing is still present. We corrected for this using the heasoft program \emph{rgsrmfsmooth} (A. Rasmussen). This program calculates a brightness profile, based on an image, in the direction of the RGS dispersion axis. It then uses this profile to adjust the response matrix to correct for the line smearing. In our analysis a Chandra image was used to create the profile, to obtain maximum precision in the correction.

The spectra were analysed using the SRON SPEX package \citep{spex}. Before the spectra were fit, we performed the SPEX optimal binning. This bins the data into optimal bins, making use of the statistics of the source as well as the instrumental resolution.

All standard RGS and MOS reduction tasks were done using XMM SAS version 10.0.0. 

\section{Results}

\subsection{Overall spectra}
\begin{figure*}
\centering
\includegraphics[width=15cm]{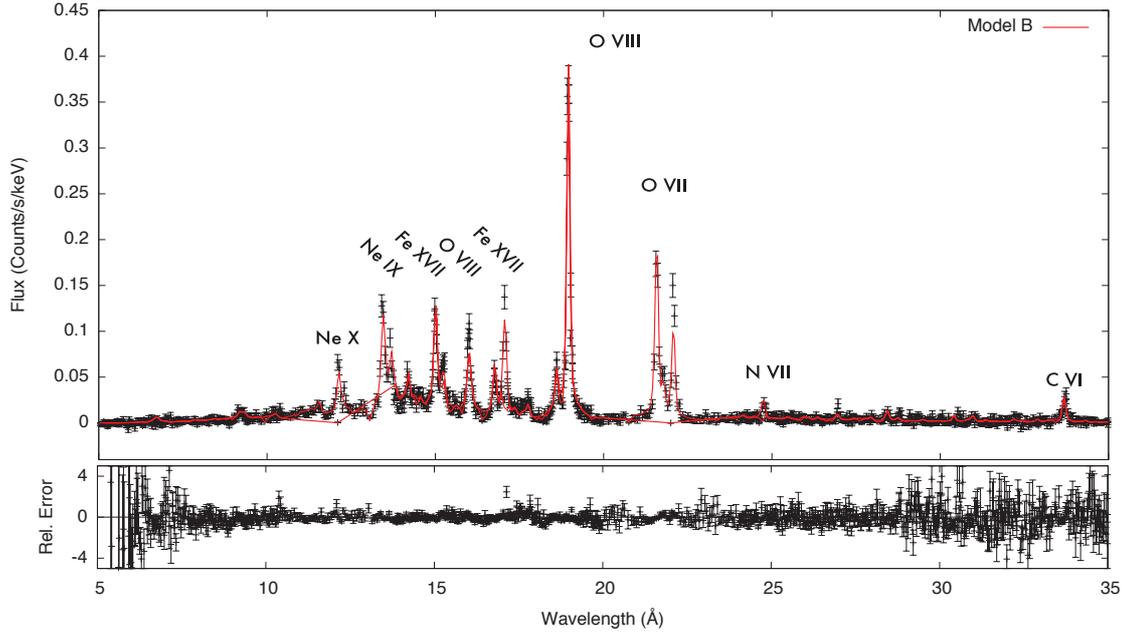}
\caption{The total 2001 RGS1 and RGS2 spectrum of SNR 0506-68 in the range 5--35\,\ang. The best fit two-component NEI model (see text) is plotted in red.}
\label{Fig.:total0101}
\end{figure*}

\begin{figure}
\centering
\resizebox{\hsize}{!}{\includegraphics{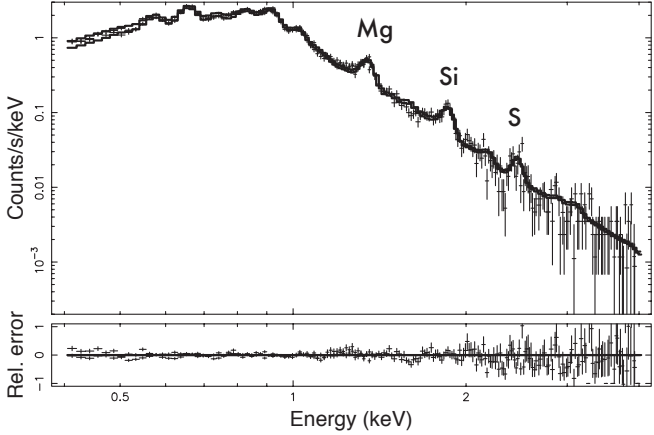}}
\caption{Plot of the MOS 1 and MOS 2 data of 2002. The model seems to fit the data well.  }
\label{Fig.:mos}
\end{figure}

The total RGS spectrum of SNR 0506-68 is shown in Fig.~\ref{Fig.:total0101} and the MOS spectrum is shown in Fig.~\ref{Fig.:mos}. As is clear from Fig. \ref{Fig.:total0101}, the spectrum is dominated by emission lines of highly ionised O, Fe, Ne, $\ion{N}{}$ and $\ion{C}{}$. The oxygen emission is particularly present with the notable $\ion{O}{vii}$ line triplet at $\sim$ 22\,\ang and the $\ion{O}{viii}$ lines at $\sim$19 and $\sim$16 \ang. The fact that both $\ion{Ne}{}$ and $\ion{C}{}$ are present suggests that there exist both a cool and hot component in the local plasma. Thus, like \cite{someyaetal2010}, the spectra were fit using two non-ionisation equilibrium (NEI) models. This model attempts to fit the data with two values of the parameters emission measure $n_{\rm\scriptsize{e}}\,n_{\rm\scriptsize{H}}\,V_{\rm\scriptsize{X}}$, electron temperature $T_{\rm e}$ and ionisation age $n_{\rm e}t$.  We used the standard SPEX absorption model to produce the hydrogen column $N_{\rm\scriptsize{H}}$ to the remnant. Before fitting with the RGS, we first obtained the high $T$ continuum slope by fitting the MOS data. When fitting the RGS data, the abundances were first fixed at the LMC / ISM abundances found by \cite{hughesetal1998}, and the ionisation parameters of the two NEI components were coupled. When the different temperatures of the models had converged, the ionisation parameters as well as some of the abundances were released to obtain a better fit. 

For the RGS data, models with different ionisation timescales $n_{\rm e}t$ and temperatures T give more or less comparable fits to the data. The models that were investigated can roughly be divided in four categories which are listed in Table \ref{tab:models}. The model listed at the bottom row of the table (model D) consists of two NEI components which are in ionisation equilibrium, i.e: $n_{\rm e}t \geq 10^{12}$ cm$^{-3}$ s. This model is comparable to the best fit model found by \cite{someyaetal2010}. A lower C-statistic was found for a two component NEI model, in which one of the components had a somewhat lower ionisation timescale (model C). This indicates that at least part of the remnant's plasma is out of ionisation equilibrium. The two best-fit models of table \ref{tab:models} are a two component NEI model, in which both components have a low $n_{\rm e}t$ (model A) and a cooling model (model B, explained below). Because the difference in C-statistic between these models is very minor, considering the amount of degrees of freedom, we listed the parameters of both the models in table \ref{tab:abundances}. The ionisation parameters of model A and B are comparable to those found by \cite{hughesetal2006}.
 
The best fit model of the 2002 RGS spectrum, the cooling model (model B), deserves special attention. This is a model in which one of the NEI components is inverted, i.e.: the initial temperature of this component is higher than the final temperature. This cooling model, with an initial temperature of 3.0 keV, reproduces especially the $\ion{O}{vii}$ resonance to forbidden line ratio better than a non-cooling two temperature NEI model. Furthermore it produces radiative continua in the higher energy parts of the spectrum, which become important in the MOS spectra. Physically, the cooling model corresponds to a plasma for which the cooling rate exceeds the recombination rate, which can cause overionisation. As mentioned in the introduction, rapidly cooling plasmas have been observed before in, among others, the mixed morphology remnants IC 443 and W49B. The fact that the overionisation model works well for SNR 0506-68 could mean that overionisation is not limited to mature SNRs of the mixed morphology class alone.  

\begin{figure}
\centering
\resizebox{\hsize}{!}{\includegraphics{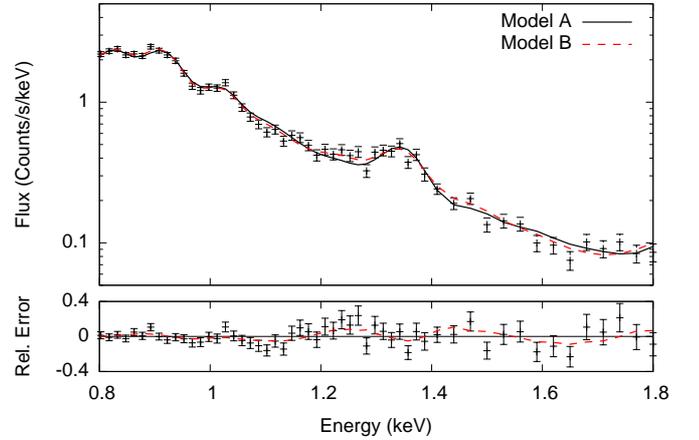}}
\caption{A plot of the mos 2 data with model A and model B in the range 0.8 - 1.8 keV. The bottom part of the figure shows the relative error between model A and the data, while the dashed red line shows model B - model A. It is clear that model B follows the overall shape of the data better than model A. For example, model A shows a residual at $\sim1.25$ keV, whereas model B improves the fit of this region due to the $\ion{Fe}{xvii}$ recombination edge at 1.26 keV.}
\label{fig:mos_res}
\end{figure}

The MOS spectra were also fit with different combinations of $n_{\rm e}t$ and $T$, similar to the RGS data. Again, model A and model B gave the best fit to the data. In contrast to the RGS data, however, the difference in best fit C-statistic between model A and B \emph{is} significant for the MOS data, namely 100\,$\sigma$. When comparing the fits of model A and B, it is not immediately clear where the difference of 100\,$\sigma$ in C-stat comes from. Distinct radiative recombination edges, as have been observed in e.g. W49B, are not clearly visible in our spectrum. Nevertheless, a more detailed inspection reveals statistically significant differences. Fig. \ref{fig:mos_res} shows the model fits in the energy range $0.8-1.8$ keV. The dashed red line shows the best fit model, model B, while the black line shows model A. The bottom part of the figure shows the residuals of the data with model A, with the difference between model B and A plotted as a dashed red line. This shows that model B follows the overall shape of the data better than model A. The presence of the Fe XVII recombination continuum at 1.26 keV, for example, improves the fit in the 1.2-1.3 keV range by lowering the C-stat by $\sim9$\,$\sigma$.  

As said, the abundances were coupled between the different model components. However, some parameters were decoupled to check if there were significant differences between the hot and cool component. The only significant difference occurred when the $\ion{Fe}{}$ abundances were decoupled. In all cases this lead to a significant improvement of the fit (the C-stat/d.o.f. decreased to 4144/2686 for our best fit model). The $\ion{Fe}{}$ abundance of the low temperature component jumped to values of five times solar in case of the RGS data, while the high temperature $\ion{Fe}{}$ abundance decreased to $\sim0.1$ solar. In addition, the temperature of the lower $T$ component decreased to a value of $\sim0.14$ keV. It is possible that there is some cool Fe present in the SNR, as this has been found before in mature SNRs \citep[e.g.][]{Uchidaetal2009}. However, at a temperature of 0.14 keV the iron emission increases considerably when the temperature is raised by even a relatively small amount ($\sim0.1$ keV). It is possible that the $\ion{Fe}{}$ emission requires a higher temperature than other elements or that there are small temperature gradients present, and that the model compensates for this by increasing the Fe abundance. Since we considered a five times solar abundance of $\ion{Fe}{}$ non-physical in SNR 0506-68, we kept the abundances between the two model components fixed. The RGS is less sensitive at higher energies, so the MOS data were used to constrain the abundances of Mg, Si and S.

\begin{table}
\renewcommand{\tabcolsep}{.2cm}
\renewcommand{\arraystretch}{1.2}
	\caption{\small RGS data C-stat / d.o.f. of the tried models for the different observations. }
		\begin{tabular}{lllll   }
	 	Model & $n_{\rm e}t_1$ (high $kT$) & $n_{\rm e}t_2$ (low $kT$)&2000 & 2002 \\
		\hline
		A&low &low &4801 / 3160 &4228 / 2684  \\
		B &low &low&4806 / 3161 &4218 / 2685  \\
		C& low &high&4820 / 3160 &4234 / 2684  \\
		D &high & high&4832 / 3159  &4282 / 2683  \\
		
		\end {tabular}
		 \tablefoot{All models consist of two NEI components which had different kT; one high (0.6-0.85 keV), and one low (0.1-0.25 keV). The different components also have different $n_{\rm e}t$, high $\sim10^{14}\,$cm$^{-3}$ s and low $\sim10^{10-11}\,$cm$^{-3}$ s. Model B is a cooling model (see text). The parameters of models A and B are listed in Table 2.}
		\label{tab:models}
\end{table}

\begin{table*}
\renewcommand{\tabcolsep}{.2cm}
\renewcommand{\arraystretch}{1.2}
   \begin{centering}
   \caption{\small Best fit parameters for the 2 NEI and the cooling model. }
 	\begin{tabular}{l@{ }lllll}
 
  & & \multicolumn{2}{c}{Model A} & 	    \multicolumn{2}{c}{Model B} \\ 
   \cline{3-4} 
   \cline{5-6}
  \multicolumn{2}{l}{Parameter} & RGS & MOS & RGS & MOS  \\ \hline
\Nh &$(10^{21}$ cm$^{-2})$& $1.14\pm0.01$ & $1.14\pm0.01$ & $1.14\pm0.01$& $1.14\pm0.01$ \\ 
$n_{\rm e}n_{\rm h}V_1$& $(10^{58}$ cm$^{-3})$ &$20.5\pm0.5 $  &$ 14.9\pm0.2$  &$10.5\pm0.3$ &$11.4\pm0.1$ \\
$n_{\rm e}n_{\rm h}V_2$&$ (10^{58}$ cm$^{-3})$ & $99.6\pm3.7$    &$ 103.5\pm1.6$  & $190\pm7.5$&$ 138.5\pm1.6$\\
Preshock $kT$ & (keV) & -& - & 3.0  & 3.0 \\
$kT_1$& (keV) &  0.85 (fixed)    &$ 0.85\pm0.01$	&0.85 (fixed) &$0.85\pm0.01$ \\
$n_{\rm e}t_1$& $(10^{10}$ cm$^{-3}$ s) &$4.67\pm0.32$  &$ 7.20_{ -0.17}^{ 0.06}$& $2.32\pm 0.13$ &$6.7\pm0.2$\\	 
$kT_2$& (keV)&$0.19\pm0.00$       &$ 0.22\pm 0.00$  &$0.15\pm0.00$	 & $0.18\pm0.00$\\
$n_{\rm e}t_2$ &$(10^{10}$ cm$^{-3}$ s) & $41.2\pm6.6$        &$ 52.1_{ -3.9}^{ 610}$ &$54.5\pm 3.7$ & $99.6\pm2.7$\\
$L_{\rm X}$ & (10$^{34}$ erg s$^{-1}$) & 12.5 & 9.4 &14  &8.8  \\
C-stat / d.o.f.\tablefootmark{a} && -  & 1170/632 & - & 1068/633

\\
 Element & \multicolumn{4}{c}{Abundance (wrt solar)}  \\
\hline	
\multicolumn{2}{l}{C}  &$0.55\pm0.08   $	  &-&					  $0.18\pm 0.03$&	 -\\
\multicolumn{2}{l}{N}  &$0.06\pm0.02  $  	  &-&					  $0.07\pm 0.02$&	 -\\
\multicolumn{2}{l}{O}  &$0.20\pm0.01  $  	  &$ 0.19\pm0.01$ &		  $0.26\pm 0.01$&	$0.22\pm0.01$ \\
\multicolumn{2}{l}{Ne} &$ 0.22\pm0.02  $ 	  &$ 0.22\pm0.02$ &		  $0.33\pm 0.03$&	$0.24\pm0.02$\\
\multicolumn{2}{l}{Mg} &-			   &$ 0.31\pm0.02$ &		 	 -		&       $0.27\pm0.02$\\ 
\multicolumn{2}{l}{Si} & -  			   &$ 0.25\pm 0.03$ &		  	-		&       $0.19\pm0.02$\\ 
\multicolumn{2}{l}{S}  &	 -		  &$ 0.33\pm 0.09$	&			  -	 	&       $0.31\pm0.07$\\ 
\multicolumn{2}{l}{Fe} &$ 0.26\pm0.01 $     &$ 0.29\pm0.01$ &		  $0.38\pm 0.02$&	$0.43\pm0.02$\\

	\end{tabular}
    \tablefoot{The \cite{abundances} Solar abundances were used.The abundances are comparable between the instruments.  	\\
    \tablefoottext{a}{The C-stat/d.o.f. of the RGS data can be found in Table 1. }
    }
    \label{tab:abundances}
\end{centering}
\end{table*}

\subsection{Detailed line spectroscopy}
\begin{figure}[!h]
\centering
\resizebox{\hsize}{!}{\includegraphics{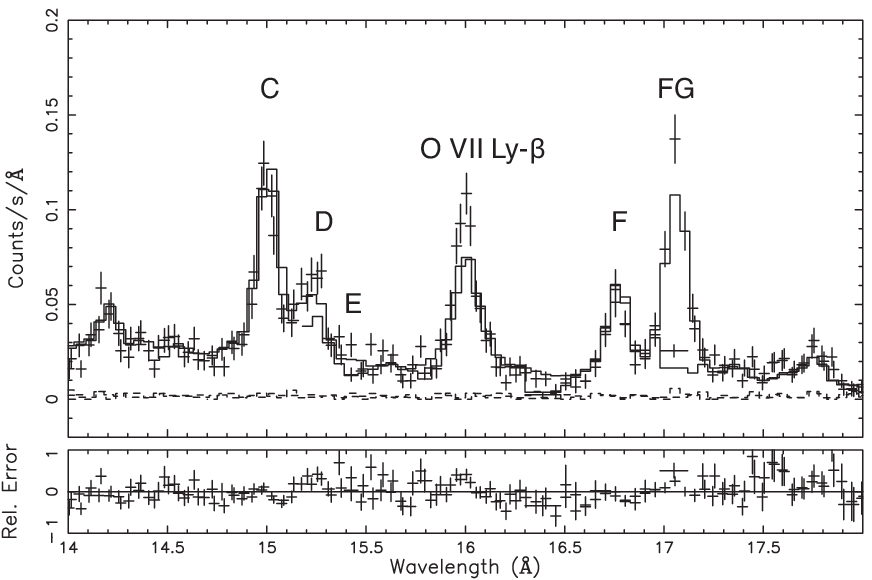}}
\caption{The 2002 RGS1 and RGS2 spectrum of SNR 0506-68 in the range 14-18 \ang. The best fit two-component NEI model (see text) is plotted in solid black. The model has trouble fitting the 15 and 17 \ang $\ion{Fe}{xvii}$ as well as the $\ion{O}{viii}$ Ly$\,\beta$/Ly$\,\alpha$ line ratio.}
\label{Fig.:1418}
\end{figure}

As shown above, model B, i.e. a cooling plasma, gives a good fit to our data. If the plasma is indeed rapidly cooling, there could be some other spectral indications. Below we will investigate several known plasma diagnostics to obtain detailed information about the spectrum.

An important diagnostic is the ratio of the different lines of the $\ion{O}{vii}$ line triplet. In the 2002 spectrum, the forbidden line at 22.08 \ang is underestimated by the non-cooling model, while the resonance line at 21.6\ang is overestimated. As mentioned above, this line triplet was best reproduced by manually tweaking an NEI model to mimic a recombining plasma. Recombination preferentially populates the triplet levels that feed the forbidden line, while the resonance line is populated mainly by collisional excitation. As such, an enhanced forbidden line suggests the presence of enhanced recombination in the plasma. This will be discussed in more detail in paragraph \ref{par:Gratio}. 

\begin{table}[!h]
   \begin{centering}
   \caption{\small Line fluxes of a number of emission lines calculated with and without hydrogen absorption. }
 	\begin{tabular}{llll}
\hline
	     line &$\lambda$ (\AA) & Intrinsic Flux\tablefootmark{a}  &Flux without absorption\tablefootmark{a}  \\
	     \hline
	
$\ion{Fe}{xvii}\,C$ &$15.01$&$1.11 \pm0.09$& $0.76 \pm0.06$ \\
$\ion{Fe}{xvii}\,D$ &$15.24$&$0.76 \pm0.09$& $0.50 \pm0.06$ \\
$\ion{Fe}{xvii}\,F$&$16.78$&$0.52 \pm0.08$& $0.32 \pm0.05$ \\
$\ion{Fe}{xvii}\,GH$ &$17.05$&$2.21 \pm0.16$& $1.22 \pm0.09$ \\
$\ion{O}{viii}\,\rm Ly\, \beta$&$16.01$&$1.00 \pm0.09$& $0.65 \pm0.06$ \\
$\ion{O}{viii}\,\rm Ly\, \alpha$&$18.97$&$7.63 \pm0.25$& $3.88 \pm0.13$ \\
$\ion{O}{vii}\,r$&$21.60$&$5.90 \pm0.41$& $2.30 \pm0.16$ \\
$\ion{O}{vii}\,i$&$21.80$&$1.37 \pm0.28$& $0.55 \pm0.10$ \\
$\ion{O}{vii}\,f$&$22.10$&$4.69 \pm0.38$& $1.69 \pm0.14$ \\
$\ion{C}{vi}\,\rm Ly\, \beta$&$28.47$&$0.48 \pm0.19$& $0.19 \pm0.06$ \\
$\ion{C}{vi}\,\rm Ly \,\alpha$&$33.74$&$5.83 \pm0.80$& $0.71 \pm0.10$ \\
	
	\end{tabular}
	\tablefoot{
	\tablefoottext{a}{In units of $10^{44}$ ph s$^{-1}$.}
	}
	    \label{tab:linestrenghts}
\end{centering}
\end{table}

All tested models have trouble fitting the $\ion{Fe}{xvii}$ 15-17 \ang line ratio (see Fig. \ref{Fig.:1418}). These lines are formed by the transitions from the 3d and 3s levels of the Ne-like ion to the ground state. We used the same labeling of the lines that was used by \cite{gillaspyetal2011}. The fact that the 17\ang line blend is stronger than the 15\ang line can be another sign of enhanced recombination \citep[]{liedahletal1990}. To obtain the $\ion{Fe}{xvii}$ line ratios, the best fit overall model was used, while the contribution to the emission by $\ion{Fe}{xvii}$ lines was excluded from this model; this ensures that contributions from, e.g., higher lines of the O VIII Ly series are taken into account. The $\ion{Fe}{xvii}$ line complex in the range of 14-18\ang was then fit with five gaussians, with a fixed \Nh of $1.14\times10^{21}\,$cm$^{-2}$. The observed line strengths are listed in table \ref{tab:linestrenghts}. We can compare our $\ion{Fe}{xvii}$ line ratios with recent laboratory measurements to obtain more information about the plasma. \citet{gillaspyetal2011} measure the 3s/C and C/D ratios at different electron beam temperatures, where 3s = F+G+H. Our observed 3s/C ratio of $2.45\pm0.2$ and C/D ratio of $1.46\pm0.13$ both correspond to an electron beam temperature of $\sim0.85$ keV, which is near the collisional excitation threshold. The observed line ratios show no indication of a high 3s/3d ratio indicative of recombination.

At 16 \ang, the $\ion{O}{viii}$ Ly\,$\beta$ line is underestimated by the model. By fitting the 15-20 \ang region with an absorbed continuum and gaussians, a Ly$\,\beta$/Ly$\,\alpha$ ratio of $0.13\pm0.01$ is obtained. This value is in agreement with CIE values found for this ratio at $T \simeq 4-5 \times 10^6$ K \citep{smithetal2001}. 

\begin{figure}[!ht]
\centering
\resizebox{\hsize}{!}{\includegraphics[angle=0]{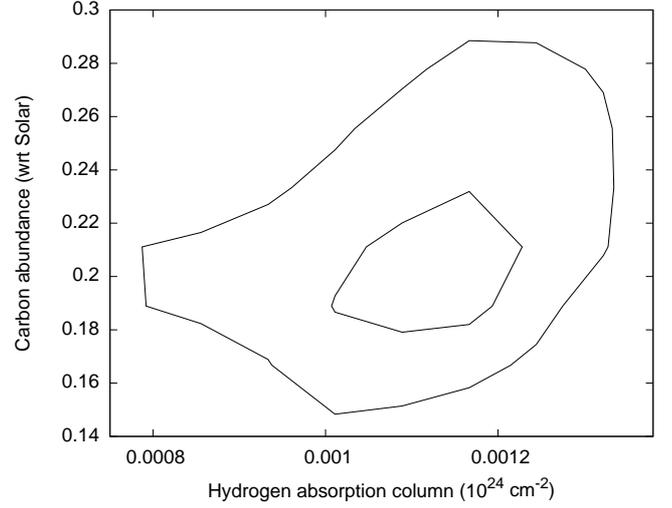}}
\caption{Contour plot of the carbon abundance and the Hydrogen absorption column made using model B. The contours represent the 1 and 2\,$\sigma$ confidence regions.  }
\label{fig:contour}
\end{figure}

$\ion{C}{vi}$ at 33.7\ang also has a strong presence and is detected at a 9\,$\sigma$ level. Because it is such a strong line, and it is affected heavily by absorption, it can be used to constrain the absorption column to the remnant. This was done by making a contour plot (Fig. \ref{fig:contour}) of the \Nh and the C abundance after a good fit was obtained. This resulted in an \Nh of $1.14\times10^{21}$ cm$^{-2}$, which was used in all our models. Note that in our model the $\ion{C}{vi}$ originates from the coolest component only (kT = 0.15-0.20 keV). At these temperatures C is mostly ionized, with the $\ion{C}{vi}$ fraction being as low as 25\%. In principle additional $\ion{C}{vi}$ emission could come from an even lower temperature component. However, since the cooling timescale for such a component is very short a major contribution does not seem very likely.

\subsection{G-ratio}
\label{par:Gratio}

\begin{figure}
\centering
\includegraphics{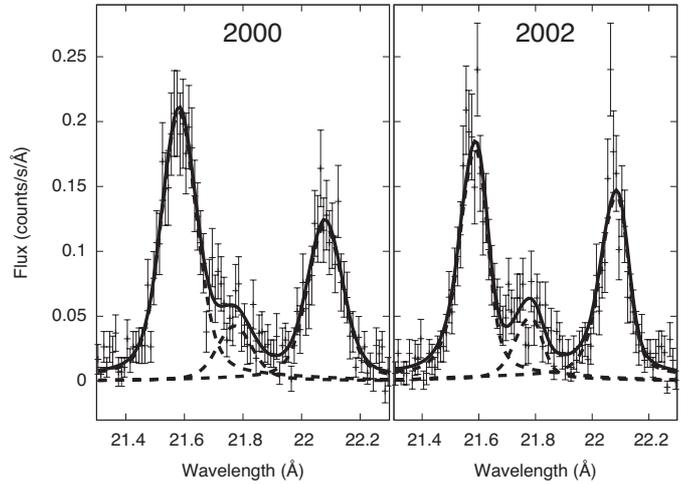}
\caption{The 2000 and 2002 observation of the OVII He-$\alpha$ triplet. It is clear that the 2002 observation has a higher effective spectral resolution.}
\label{Fig.:ovii}
\end{figure}

Fig. \ref{Fig.:ovii} shows the 2000 and 2002 observations of the $\ion{O}{vii}$ He-$\alpha$ triplet. This triplet consists of a resonance, forbidden and intercombination line. F ($\lambda$ = 22.098) $1s2s^3 S_1 \rightarrow 1s^{2}~ ^{1}S_0$ is the forbidden transition, I ($\lambda$ = 21.804, 21.801) is the sum of the two intercombination transitions $1s2 p^3 P_{1,2} \rightarrow 1s^{2}~ ^{1}S_0$, and R ($\lambda$ = 21.602) is the resonance transition $1s2p ^{1}P_1\rightarrow 1s^{2}~ ^{1}S_0 $. Both the 2000 and the 2002 observation are well-fit by three gaussians with a fixed \Nh. There are some differences between the two observations, the most notable one being that the F/R ratio is larger in the 2002 observation. In addition, the lines in the 2000 observation are broader, which is an effect of the orientation of the RGS dispersion axis. The total flux in the line triplet is approximately equal between the observations. 

\begin{figure}
\centering
\resizebox{\hsize}{!}{\includegraphics[angle=-90]{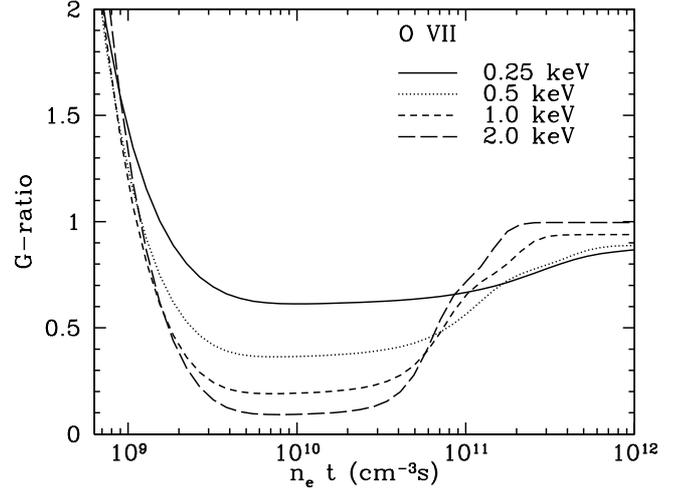}}

\caption{G-ratio of $\ion{O}{vii}$ at different temperatures, as a function of ionisation parameter. These curves where calculated using SPEX. }
\label{fig:gratios}
\end{figure}

An interesting quantity which can be derived from these lines is the so-called G-ratio; $G \equiv (F + I)/R$ \citep[e.g.:][]{porquetetal2011}. This quantity equals $0.87\pm0.09$ for 2000 and $1.19\pm0.09$ for 2002, giving a combined ratio of $0.99\pm0.06$. As the 2002 observation has a higher effective resolution, the G-ratio of that observation may be more reliable. Fig. \ref{fig:gratios} shows a plot of G-ratios for different n$_{\rm e}t$. The values of the G-ratio for different temperatures reach a constant value as the plasma approaches collisional ionisation equilibrium (CIE). If we take the mean value of the above G-ratios, the calculated G-ratio at high $n_{\rm e}t$ and $kT=0.2$ keV lies just within the error bars and is thus as expected. If the 2002 observation is indeed more reliable, the deduced G-ratio of the plasma lies above this CIE value and the plasma is over-ionised and could be recombining. In principle the excess photons present in the forbidden line should show up as a recombination edge at 16.78\ang. A recombination edge in addition to the already present continuum was not found, however.  

\section{Discussion}

We made a detailed spectral analysis of the SNR 0506-68 using mainly the RGS instrument aboard the XMM-Newton telescope. The best fit to the overall mos and RGS spectra is model B:  a two component NEI model, of which one component is inverted. We investigated the hypothesis that the plasma is cooling leading to an overionisation, using some known plasma diagnostics. Of those diagnostics, only the 2002 $\ion{O}{vii}$ triplet line ratio confirms the hypothesis. There are however, several other physical mechanisms which can cause the observed line ratio. 

Resonance scattering can cause photons of resonance lines to be scattered in the direction of least optical depth, reducing the line flux if the optical depth along our line of sight is high. This process was suggested to be responsible for the observed $\ion{O}{vii}$ F/R ratio in the SNR DEM L71 \citep{kurtetal2003}. The scattered photons are not lost, however, so the remnant must have a specific geometry, i.e. it cannot be spherically symmetric, or resonance scattering will have no effect whatsoever. The optical depth in the $\ion{O}{vii}$ line equals about 6 in our line of sight, without taking microturbulent velocity into account. The optical depth equals 1, however, at a turbulent velocity of 80 km/s and decreases even more at higher values. A high optical depth could result in a significant reduction of the resonance line flux  \citep{kaastramewe1995}, which means that resonance scattering could be significant in this remnant. 

Charge exchange occurs when a highly ionised gas enters a neutral gas region, such as a molecular cloud. H-like ions collide with the neutral ions to form excited He-like ions which decay by a radiative cascade. This process enhances the forbidden line due to the higher statistical weight. There have been no reports of neutral gas regions close to the remnant, so the probability of the charge exchange scenario occurring remains uncertain.

\subsection{Cooling}

As mentioned above, recombination can influence the observed line ratios. For recombination to dominate, the plasma must be over-ionised and thus must have cooled faster than the recombination process happens. We now investigate if SNR 0506-68 can cool fast enough for this to occur. 

The expanding plasma cools in two ways: through radiation and due to adiabatic expansion. A timescale for radiative cooling can be obtained by dividing the internal energy of a gas by the luminosity:
\begin{equation}
\label{eq:1}
\tau_{\rm rad} \simeq \frac{\frac{3}{2}(n_{\rm e}+n_{\rm H}+n_{\rm He})kT}{n_{\rm H}^2\Lambda} = \frac{3kT}{n_{\rm H}\Lambda},
\end{equation}

where $\Lambda$ is the cooling rate in units of erg cm$^3$ s$^{-1}$, $k$ is Boltzmann's constant, and $n_{\rm H}$ is the hydrogen/proton number density. 

For an adiabatic gas, PV$^\gamma$ = constant. By the same token, TV$^{\gamma-1}$ = constant. The following formula for the adiabatic cooling timescale can thus be derived for a spherical volume:
\begin{equation}
\tau_{\rm ad} = \left(-\frac{T}{\dot{T}}\right)_{\rm ad} =  \left(\frac{1}{\gamma-1}\right)\frac{R}{3\dot{R}} \,.
\label{eq:2}
\end{equation}  

A total cooling timescale as a result of radiative and adiabatic cooling is now given by $\tau_{\rm cool}^{-1}=\tau_{\rm rad}^{-1}+\tau_{\rm ad}^{-1}$. When a plasma cools, the radiative cooling rate of the remnant is proportional to the density squared. The recombination rate also increases proportional to the density squared. If adiabatic cooling is not important compared to radiative cooling, the remnant will stay in ionisation equilibrium. Radiative cooling is faster than recombination below $10^6$ K or if the elemental abundances are strongly enhanced. For an over-ionised plasma, it is required that $\tau_{\rm rec} > \tau_{\rm cool}$. The recombination timescale for $\ion{O}{viii} \rightarrow \ion{O}{vii}$ in the range $10^6 \leq T \leq 10^7$ is approximately equal to $(1.3\times10^{-12}\,n_{\rm e})^{-1}$. Using \ref{eq:1}, \ref{eq:2} with $\gamma=5/3$, this leads to:
\begin{equation}
\label{eq:3}
\frac{2\dot{R}}{R} + \frac{n\Lambda}{3kT} < \frac{1}{1.3\times10^{-12}n_{\rm e}}\,.
\end{equation}

Note that the remnant needs to have reached ionisation equilibrium before adiabatic cooling causes an over-ionised plasma. For $R \simeq 10$ pc, $\dot{R} \simeq 400$ km s$^{-1}$, $\Lambda = 3\times10^{-23}$ erg cm$^3$ s$^{-1}$, corresponding to a 0.2 keV plasma at LMC abundances \citep[][]{klaartjecooling} and a density $n$ = 10 cm$^{-3}$, the recombination timescale (using the above formula) is about five times lower than the cooling timescale, and depends strongly on the density. 

To expand the above equation to a larger temperature range, we can rewrite $R/\dot{R} = \beta^{-1} t$, where $\beta$ is the expansion parameter and $t$ is the age of the remnant. Values of $\beta$ vary between about 0.4 for a remnant in the Sedov expansion phase and 0.25 in the snowplough phase. In addition, we can use the temperature dependent $\ion{O}{viii} \rightarrow \ion{O}{vii}$ (radiative + dielectronic) recombination rates from \cite{shullvansteenberg1982} as well as the complete $\ion{O}{}$ radiative cooling curves from \citet{klaartjecooling}.

\begin{figure}
\centering
\resizebox{\hsize}{!}{\includegraphics[angle=0]{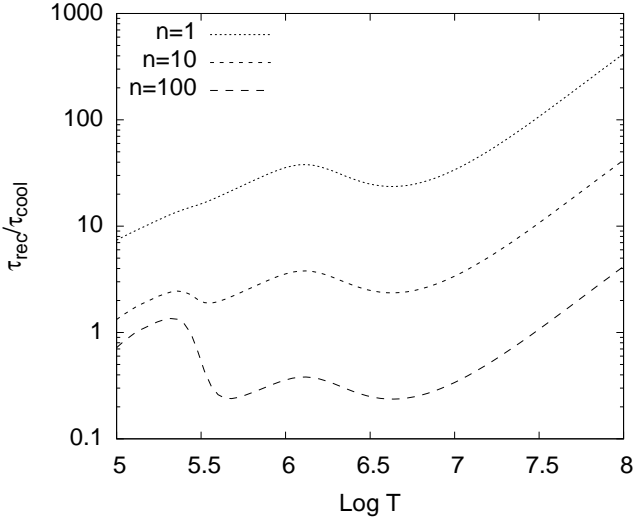}}

\caption{A comparison of the oxygen recombination timescale with the cooling timescale at different densities. Low density favours the appearance of an overionised plasma, because the recombination timescale lowers proportional to the density, as does radiative cooling, but radiative cooling only becomes important at log $T \leq 10^6\,$K. These curves where calculated using LMC abundances. It should be noted that a plasma at LMC abundances is less likely to reach an over-ionised state at $T<10^6$ K, because of the lower radiative cooling rate.}
\label{fig:tadtrad}
\end{figure}

Fig. \ref{fig:tadtrad} shows this relation between the cooling and the recombination time using $t = 10^{11}$ s and $\beta=0.4$. The behaviour at high temperature is dominated by the adiabatic cooling, while radiative cooling becomes important at $T < 10^6$ K. An increased density increases the recombination rate, which decreases the likelihood of plasma in SNRs to become over-ionised at temperatures at which radiative cooling is unimportant. It should be noted that the above derivation is a first order estimate. We can however still use it to check if the conditions in SNR 0506-68 are likely to cause an over-ionised plasma. At a $T\sim10^6$ K and $n \sim 10$ cm$^{-3}$ the cooling timescale is about two times lower than the recombination timescale. This is a discrepancy with the above calculation, so we can conclude that either our remnant is not in the Sedov stage of evolution (i.e. $\beta<0.4$), or that the shock velocity or density estimations are not right. 

We can check if the overionisation observed in other SNRs can be explained using the above relation. In W49B, recombining $\ion{Fe}{xxv-xxvi}$ was found. Using their respective recombination and radiative cooling rates at $T =  1.5$ keV and $n = 10$ cm$^{-3}$, the cooling timescale is indeed lower than the recombination timescale.

Note that the above derivation for the cooling rate is somewhat conservative, as we did not take into account dust cooling. This is in particular true for SNR 0506-68, for which \citet{williamsetal2006} found an infrared luminosity of $8.7\times10^{36}$ erg s$^{-1}$. This is about 60 times higher than the observed X-ray luminosity. In general dust cooling increases the cooling rate of the plasma, which makes it easier to reach an over-ionised state. 

\subsection{Explosion parameters}

The data obtained from fitting allow us to constrain several interesting parameters. For this we need the volume of the plasma, which we obtained using the Chandra data. The radius of the sphere in which the plasma is contained is about 38'', which corresponds to a radius of 9.1\,pc at a distance of 50\,kpc. We therefore estimate the volume of the sphere to be $9 \times 10^{58}$ cm$^3$. The fact that the radiation is distributed anisotropically over the remnant as well as the fact that the emission does not come from a purely spherical region, can be accounted for by using a filling factor f. A filling factor within 10-25$\%$ seems reasonable. 

Using the normalisation factor $n_{\rm\scriptsize{e}}n_{\rm\scriptsize{H}}V_{\rm\scriptsize{X}}$ of the best fit model to the RGS spectra, with \nel = \nh$/1.2$  and $V_X$ the volume, we obtain a density \nh$=11(f_{0.1})^{-0.5}$ cm$^{-3}$ and \nh$=5(f_{0.1})^{-0.5}$ cm$^{-3}$ for respectively the cool and the hot component of the non-cooling NEI model, where $f_{0.1}$ represents a filling factor of 10$\%$. The density of the cool component of the cooling model is slightly higher at $14(f_{0.1})^{-0.5}$. The density of the low T component is somewhat higher than previously found values. \cite{hughesetal2006} find densities of 10-23 cm$^{-3}$ using X-ray data, \cite{dickelmilne1998} used radio data to find a maximum density of 10 cm$^{-3}$ and \cite{williamsetal2006} found a density of 5.8 cm$^{-3}$ by modelling dust emission in the remnant. These are mean densities over the whole remnant and it is probable that at some locations the densities are higher. 

With these densities and the ionisation parameters the shock ages of the cool and hot components are $\sim1200$ and $\sim115$ yr. Different parts of the remnant were thus shocked at different times, which suggests that there are some density fluctuations in the surrounding ISM, which is also clear from Fig. 1. It has already been hinted by \cite{hughesetal2006} that the open cluster HS 114 \citep{hodgesexton1966}, which lies on the brighter, high density side of the remnant, may be the cause of the observed brightness gradient. 

The total swept up mass is given by $M_{\rm{swept}} \sim n_{\rm H}m_{\rm p}fV$, where $m_{\rm p}$ is the proton mass. $M_{\rm{swept}} \sim 88 \sqrt{f_{0.1}}$ \Msun for the cool component and $M_{\rm{swept}} = 37\sqrt{f_{0.1}}$ \Msun for the hot component.

Due to the brightness gradient and the fact that a low and high temperature NEI model are necessary to fit the spectrum of the remnant, an age estimation is difficult. Using a Sedov model ($t_{\rm Sedov}=4.3\times10^2\left(\frac{R}{1\, \rm{pc}}\right)\left(\frac{kT_{\rm e}}{1\, \rm{keV}}\right)^{-0.5}$), the age of the cool component can be estimated at $\sim9000$ year, while the age of the hot component equals $\sim4000$ year. A Sedov model assumes that the remnant is expanding in a homogeneous medium which, judging from the anisotropic emission, is not quite valid in the case of SNR 0506-68. Since the hot, less dense component of the plasma is likely to have expanded relatively undisturbed, the age estimation of this component is probably more accurate. 

\section{Conclusions}
\begin{itemize}
\item The SNR 0506-68 is best fit by a two component NEI model, of which one of the components is cooling. This suggests that the SNR may be cooling, but the emission line diagnostics marginally agree with this scenario. 
\item The most natural explanation for the enhanced G-ratio of the $\ion{O}{vii}$ He-$\alpha$ is that part of the plasma is recombining, although resonance scattering cannot be ruled out. 
\item When an expanding plasma reaches near-ionisation equilibrium, adiabatic cooling can cause it to become over-ionised. 
\item The abundances of SNR 0506-68 are very similar to the mean LMC abundances. This, coupled with the fact that $M_{\rm swept}$ is fairly high, confirms that the emission is dominated by the emission from the shock heated ISM. 
\item Overionisation may be more widespread in mature SNRs than usually thought. 
\item The age of the remnant is uncertain and model dependent. However our models and calculations favour the lower age estimate mentioned in the literature, namely $\sim 4000$ year. 
\end{itemize}
\begin{acknowledgements}
The authors thank the anonymous referee for their positive and useful comments. We would like to acknowledge D. Kosenko and T. van Werkhoven for useful input. S.B. and SRON are supported financially by NWO, the Netherlands Organisation for Scientific Research. The results presented are based on observations obtained with XMM-Newton, an ESA science mission with instruments and contributions directly funded by ESA Member States and the USA (NASA).
\end{acknowledgements}

\bibliographystyle{aa.bst}
\bibliography{sjors}

\end{document}